\documentclass{edp-conf}
\usepackage{graphicx}
\begin{document}

\TitreGlobal{SF2A 2005}

\title{B \& Be stars in the Magellanic Clouds: rotation, evolution and variability.}
\author{Martayan, C.}\address{GEPI, Observatoire de Paris, France}
\author{Fr\'emat, Y.}\address{Observatoire royal de Belgique, Belgium}
\author{Floquet, M.$^1$}
\author{Hubert, A.-M.$^1$}
\author{Zorec, J.}\address{Institut d'Astrophysique de Paris, France}
\author{Mekkas, M.$^1$}
\runningtitle{B \& Be in the Magellanic Clouds.}
\setcounter{page}{237}
\index{Martayan, C.}
\index{Fr\'emat, Y.}
\index{Floquet, M.}
\index{Hubert, A.-M.}
\index{Zorec, J.}
\index{Mekkas, M.}


\maketitle
\begin{abstract} 
Thanks to observations at the ESO VLT with the GIRAFFE multifibers spectrograph, we have obtained spectra of 177 and 346
B-type stars in the Large and Small Magellanic Clouds respectively. We have discovered 25 and 90 new Be stars among the 47
and 131 Be stars observed in the LMC and SMC respectively. We have determined the fundamental parameters of these stars
and examined the effect of the metallicity, star formation conditions and evolution on the behaviour of the rotational
velocities. We discovered that only the future B-type stars with a strong rotational velocity in ZAMS will become a Be
star. We have also discovered short term variability in several SMC Be stars thanks to a cross-correlation with the
MACHO database. 
\end{abstract}
%
\section{Rotation, star formation conditions, metallicity and evolution in B \& Be stars.}
\begin{figure}[h]
   \centering
   \includegraphics[width=4cm,height=9cm, angle=-90]{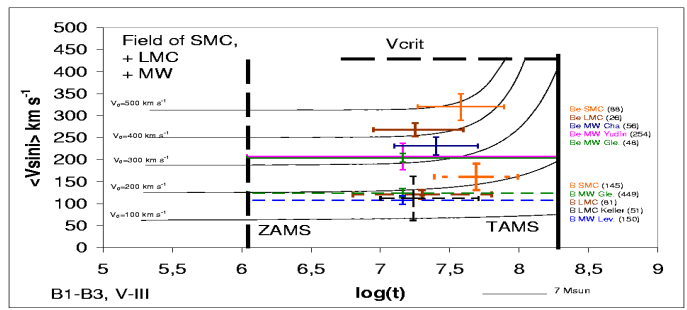}
      \caption{Comparisons of the projected rotational velcoties by spectral type selection in the MW, LMC and SMC for B
      \& Be stars.}
       \label{martayanfig1}
   \end{figure}

We have determined the fundamental parameters (effective temperature, surface gravity and projected rotational velocity)
with the GIRFIT code developped by Fr\'emat et al. (2005a) for the 523 stars observed in the Magellanic Clouds with the
VLT-GIRAFFE facilities. The main subtypes of our sample range mainly from B1 to B3. To study the influence of metallicity
on the rotational velocities, we have selected from the litterature samples of B \& Be stars in the Milky Way (MW)
(Glebocki \& Stawikowski 2000; Levato \& Grosso 2004; Yudin 2001; Chauville et al. 2001) and in the LMC (Keller 2004)
corresponding to our samples in the MC in types, classes end age if it is known. In Fig.~\ref{martayanfig1}, we show
the comparison of the projected rotational velocities for B \& Be stars in the SMC, LMC and MW (by order of decreasing
metallicity) and we show the result of the calculations of the evolution of the rotational velocity for different initial
rotational velocity for a 7 solar mass star. Then the results are the following: only the future B type stars with a
strong rotational velocity at the ZAMS will become Be star. There is an obvious effect of the metallicity on the
rotational velocity for the Be stars: the lower the metallicity is the higher the rotational velocities are. 
As Be stars are not critical rotators, the strong rotational velocity cannot explain alone the Be phenomenon, then an
additional effect is required to explain these objects such as beating of non radial pulsations or angular momentum
transport by magnetic field from core to surface. All these results are developed in Martayan et al. (2005b).
\section{Photometric variability in Be stars in the SMC.}
Among our sample, 134 Be stars have been observed by MACHO. For 13 of these stars, we have found a short-term photometric
variability (P$\leq$2.5d) with amplitude greater than 0.1 magnitude. The variabilities found are similar in amplitude and 
periods to short term variations found in the galactic Be stars (Hubert \& Floquet 1998). As in the MW, the light
modulation may be linked with stellar activity or non radial pulsations.

\end{document}